\documentclass[12pt]{article}
\usepackage[centertags]{amsmath}
\usepackage{amsfonts}
\usepackage{amssymb}
\usepackage{amsthm}
\usepackage{cite}
\usepackage{graphicx}
\usepackage{etoolbox}

\newcommand{\beq}{\begin{equation}}
\newcommand{\eeq}[1]{\label{#1}\end{equation}}
\newcommand{\bea}{\begin{eqnarray}}
\newcommand{\eea}[1]{\label{#1}\end{eqnarray}}

\patchcmd{\subequations}{}%
{}{}{}

\def\a{\alpha}
\def\b{\beta}
\def\g{\gamma}
\def\G{\Gamma}
\def\d{\delta}
\def\D{\Delta}
\def\e{\epsilon}
\def\ve{\varepsilon}

\def\h{\eta}

\def\k{\kappa}
\def\l{\lambda}
\def\L{\Lambda}
\def\m{\mu}
\def\n{\nu}

\def\r{\rho}

\def\s{\sigma}

\def\de{\partial}
\def\nb{\nabla}
\def\bbox{\de^2}

\begin{document}
\setlength{\topmargin}{-1cm} \setlength{\oddsidemargin}{0cm}
\setlength{\evensidemargin}{0cm}

\begin{titlepage}
\begin{center}
{\Large \bf Gravitational Properties of the Proca Field}

\vspace{20pt}

{\large Talal Ahmed Chowdhury,$^a$ Rakibur Rahman$^{\,a,b}$ and Zulfiqar Ali Sabuj$^{\,c}$}

\vspace{12pt}
$^a$ Department of Physics, University of Dhaka, Dhaka 1000, Bangladesh\\
\vspace{6pt}
$^b$ Max-Planck-Institut f\"ur Gravitationsphysik (Albert-Einstein-Institut)\\
     Am M\"uhlenberg 1, D-14476 Potsdam-Golm, Germany\\
\vspace{6pt}
$^c$ Department of Theoretical Physics, University of Dhaka, Dhaka 1000, Bangladesh\\

\end{center}
\vspace{20pt}

\begin{abstract}

We study various properties of a Proca field coupled to gravity through minimal and quadrupole interactions, described
by a two-parameter family of Lagrangians. St\"uckelberg decomposition of the effective theory spells out its model-dependent
ultraviolet cutoff, parametrically larger than the Proca mass. We present pp-wave solutions that the model admits, consider
linear fluctuations on such backgrounds, and thereby constrain the parameter space of the theory by requiring null-energy
condition and the absence of negative time delays in high-energy scattering. We briefly discuss the positivity
constraints$-$derived from unitarity and analyticity of scattering amplitudes$-$that become ineffective in this regard.

\end{abstract}

\end{titlepage}

\newpage
\section{Introduction}\label{sec:intro}

Fundamental particles relevant to our world must interact with gravity because of its universal nature. For gravitational interactions in flat space,
massive fields of arbitrary spin are immune from such severe restrictions as afflict their massless counterparts (see for example~\cite{Rahman:2015pzl}
for a review). Indeed massive particles do couple to gravity, in particular at the cubic level, giving rise to nontrivial gravitational form factors.
To be specific, a massive particle of spin $s\geq1$ may possess as many as $2s$ mass multipole moments\footnote{This count follows from considering
the matrix element of the stress-energy tensor between two spin-$s$ states~\cite{Rahal:1990gs}. The multipole expansion of the time-time component in terms
of spherical tensors, for example, contains $(2s+1)$ nontrivial pieces, whereas a mass dipole moment is not physically meaningful~\cite{GR-SUSY}.}.
Intuitively, what makes even a fundamental particle behave like an extended object sensitive to tidal forces is its non-zero Compton wavelength,
which sets an intrinsic size.

When it comes to massive particles of low spin, such as the Proca field, there seem to be no issues with coupling to gravity. Minimal coupling does
not lead to pathologies like Velo-Zwanziger acausality~\cite{Rahman:2015pzl}. Neither do non-minimal couplings, consistent with the symmetries of the theory,
pose any obvious problems. Of course, the theory will have a cutoff, which$-$depending on the model$-$may be parametrically smaller than the Planck scale.
Can all such effective field theories be embedded in weakly coupled ultraviolet completions? The answer is expected to be in the negative,
from arguments involving unitarity and analyticity of scattering amplitudes~\cite{Adams:2006sv} or time delays in high-energy scattering~\cite{Camanho:2014apa}.

In this article we will consider the gravitational interactions of a massive vector field. The Einstein-Proca theory~\cite{Tauber:1969ba} and its ghost-free
generalizations~\cite{Tasinato:2014eka,Heisenberg:2014rta,Hull:2015uwa,Allys:2015sht,Jimenez:2016isa,Heisenberg:2016eld} have generated a lot of interest in
recent years, especially because they provide with an attractive framework for
cosmology~\cite{Jimenez:2009ai,Tasinato:2014mia,DeFelice:2016yws,DeFelice:2016uil,Nakamura:2017dnf} and
astrophysics~\cite{Herdeiro:2016tmi,Fan:2016jnz,Minamitsuji:2016ydr,Babichev:2017rti,Chagoya:2016aar,Chagoya:2017fyl,Heisenberg:2017xda,Heisenberg:2017hwb}
(see~\cite{Heisenberg:2018vsk} for a recent review).
In this context, one may investigate the Proca self interactions~\cite{Bonifacio:2016wcb,deRham:2018qqo} or analyze the quantum
aspects~\cite{Charmchi:2015ggf,Buchbinder:2017zaa}. Sidestepping these interesting directions, we will study the effective field theory
of generalized Einstein-Proca actions containing only up to quadratic terms in the Proca field. Explicitly, we will consider the following two-parameter
family of Lagrangians:
\beq \mathcal{L}=\sqrt{-g}\left[\,\tfrac{1}{2}M_P^2 R-\tfrac{1}{4}F_{\m\n}^2-\tfrac{1}{2}m^2A^2-\tfrac{1}{2}\a G_{\m\n}A^\m A^\n+\tfrac{1}{4}\b\L^{\!-2}
L_{\m\n\r\s}F^{\m\n}F^{\r\s}\,\right],\eeq{L0}
which is an effective field theory of a real Proca field $A_\m$ coupled to Einstein gravity that contains dimensionless parameters $\a$ and $\b$, and has an
ultraviolet cutoff $\L$, where $F_{\m\n}=\nb_\m A_\n-\nb_\n A_\m$ is the Faraday tensor, $G_{\m\n}$ the Einstein tensor, and $L_{\m\n\r\s}$ the double dual
of the Riemann tensor:
\beq L^{\m\n}{}_{\r\s}\,=\,\tfrac{1}{4}\ve^{\m\n\a\b}\ve_{\r\s\g\d}R_{\a\b}{}^{\g\d}\,=\,-R^{\m\n}{}_{\r\s}+4\,\d^{[\m}_{[\r}R^{\n]}{}_{\s]}
-R\,\d^\m_{[\r}\d^\n_{\s]}.\eeq{L0.1}

The particular choice of Eq.~(\ref{L0}) corresponds to the most general Lagrangian with non-minimal couplings bilinear in the vector field and linear in
curvature, such that higher-order derivative terms are absent in the equations of motion~\cite{deRham:2011by,Jimenez:2013qsa,Allys:2015sht}.
Moreover, when expanded around flat space, this Lagrangian will contain all possible
Proca-graviton-Proca cubic couplings. Note that there are only two such nontrivial couplings: one with at most two derivatives and another with
four derivatives~\cite{Metsaev:2005ar,Metsaev:2012uy}. The first one is encoded by minimal coupling, whereas the second by non-minimal
quadrupole coupling to the Riemann tensor\footnote{Let us recall that a massive spin-1 field may possess only
monopole (mass) and quadrupole moments.}. Non-minimal couplings to the Ricci tensor and the scalar curvature do appear in our action~(\ref{L0})
but they result in trivial cubic interactions in flat space. Yet the inclusion of such terms is well justified. While the
$R_{\m\n}A^\m A^\n$-term is a natural consequence of the ambiguity in minimal coupling prescription since covariant derivatives do not commute,
the other terms are essential for having second-order equations of motion in curved space~\cite{Allys:2015sht}, thanks to the transversality
properties: $\nb^\m G_{\m\n}=0,~\nb^\m L_{\m\n\r\s}=0$.

We would like to study various aspects of the effective field theory described by the generalized Proca model~(\ref{L0}). While it is natural
to have $\mathcal{O}(1)$ values of $\a$ and $\b$, small values of $\a$ and $m/M_P$ are technically natural~\cite{tHooft:1979rat} since there is a $U(1)$
symmetry enhancement when these parameters are zero. We will therefore assume that
\beq |\a|\lesssim\mathcal{O}(1),\qquad |\b|\sim\mathcal{O}(1).\eeq{natural}
We will see that there appear specific constraints on the dimensionless couplings $\a$ and $\b$, essentially from requiring that the theory have a
weakly coupled ultraviolet completion.

The organization of this article is as follows. The rest of this section clarifies our metric and curvature conventions. Section~\ref{sec:cutoff}
employs the St\"uckelberg formalism to systematically study the degree of singularity in our model in the limit of vanishing Proca mass. This
gives a model-dependent effective field theory cutoff. Carried out in Section~\ref{sec:shockwave}, a shock-wave analysis spells out constraints on
$\a$ and $\b$ that rid our model of negative time delays in high-energy scattering events. Section~\ref{sec:forward} uses unitarity and analyticity
properties of scattering amplitudes to derive positivity constraints, which turn out to be ineffective. Summary of our results and some
concluding remarks appear in Section~\ref{sec:remarks}.

{\bf Conventions:} Our metric signature is $(- + + +)$. We use unit normalization for the totally symmetric expression $(\m_1\cdots\m_n)$ as well as
the totally antisymmetric expression $[\m_1\cdots\m_n]$ in the indices $\m_1,\ldots,\m_n$. The curvature conventions are:
\bea &\G^\r_{\m\n}=\tfrac{1}{2}g^{\r\s}\left(-\de_\s g_{\m\n}+\de_\m g_{\n\s}+\de_\n g_{\m\s}\right),&\nonumber\\
&R^\r{}_{\s\m\n}=\de_\m\G^\r_{\s\n}-\de_\n\G^\r_{\s\m}+\G^\r_{\l\m}\G^\l_{\s\n}-\G^\r_{\l\n}\G^\l_{\s\m}\,,\qquad R_{\m\n}=R^\r{}_{\m\r\n}\,,&\nonumber\eea{xxxxxx}
so that $[\nb_\m,\nb_\n]A^\r=R^\r_{~\s\m\n}A^\s$. The Levi-Civita symbol is normalized as $\ve_{0123}=+1$.

\newpage
\section{Effective Field Theory Cutoff}\label{sec:cutoff}

Einstein gravity is understood as an effective field theory valid up to the Planck scale $M_P$. When a single particle of spin $s\geq1$ and non-zero mass $m\ll M_P$ is coupled
to it, the resulting theory has a cutoff no higher than $\left(m^{2s-2}M_P\right)^{1/(2s-1)}$, as conjectured in~\cite{Rahman:2009zz,Rahman:2011ik,Kurt}.
Then, for the Einstein-Proca
theory the model independent cutoff upper bound should simply be $M_P$. This is indeed true, as we will see shortly. However, depending on the model, the cutoff may actually be
much lower than the Planck scale.

In this section we will find the cutoff scale $\L$ of the effective field theory described by the generalized Proca model~(\ref{L0}). Note that, by assumption, the presence/absence
of the quadrupole term does not affect the cutoff scale. One may therefore forgo the quadrupole coupling and work with Lagrangian~(\ref{L0}) with $\b$ set to zero. The model-dependent
cutoff will then be a function of the Proca mass $m$, the Planck scale $M_P$ and the dimensionless parameter $\a$, but \emph{not} of $\b$. The latter parameter is assumed to be
$\mathcal{O}(1)$, and so there is no contradiction since the cutoff is not sharply defined anyway. It is however reassuring to consider the quadrupole coupling term to see that
its inclusion in the Lagrangian has no bearing on the cutoff analysis.

\subsection{St\"uckelberg Decomposition}

In order to study the degree of singularity in our model in the limit of vanishing Proca mass, we take recourse to the St\"uckelberg formalism. Let us introduce a compensator
scalar field $\phi$ in the action~(\ref{L0}) by the substitution:
\beq A_\mu=B_\m-m^{\!-1}\de_\m \phi.\eeq{St1}
Then the Lagrangian enjoys, along with diffeomorphism, the following gauge symmetry:
\beq \d B_\m=\de_\m\l,\qquad\d\phi=m\l,\eeq{St2}
where $\l$ is a scalar gauge parameter. Known as St\"uckelberg invariance, the symmetry~(\ref{St2}) is just an artifact; one can always choose the unitary gauge to set $\phi=0$, and write
the Lagrangian in the original form~(\ref{L0}). However, the redundancy is useful in that one can choose a different gauge, in which the St\"uckelberg fields, i.e.,  the vector mode $B_\m$
and the scalar mode $\phi$, acquire canonical kinetic terms~\cite{Porrati:2008gv,Porrati:2008an,Porrati:2008ha}. This renders the corresponding propagators smooth in the massless limit.
To be specific, after the substitution~(\ref{St1}) has been made in the Lagrangian~(\ref{L0}), one can add the gauge-fixing term:
\beq \Delta\mathcal{L}=-\tfrac{1}{2}\sqrt{-g}\left(\nb_\m B^\m - m\phi\right)^2.\eeq{St3}
The result is the following gauge-fixed Lagrangian:
\beq \mathcal{L}_{\text{g.f.}}\equiv\mathcal{L}+\D\mathcal{L}=\mathcal{L}_{\text{EH}}+\mathcal{L}_0+\mathcal{L}_1+\mathcal{L}',\eeq{St4}
where $\mathcal{L}_{\text{EH}}$ is the Einstein-Hilbert part, $\mathcal{L}_0$ contains the scalar kinetic and mass terms:
\beq \mathcal{L}_0=-\tfrac{1}{2}\sqrt{-g}\left(g^{\m\n}\de_\m\phi\de_\n\phi+m^2\phi^2\right),\eeq{St31}
$\mathcal{L}_1$ incorporates the kinetic, mass and quadrupole-coupling terms of the vector mode:
\beq \mathcal{L}_1=-\tfrac{1}{2}\sqrt{-g}g^{\m\n}\left(g^{\r\s}\nb_\m B_\r\nb_\n B_\s+m^2B_\m B_\n\right)-\b\L^{\!-2}\sqrt{-g}\,
R^{\m\r\n\s}\nb_\m B_\r \nb_\n B_\s,\eeq{St32}
and $\mathcal{L}'$ contains other non-minimal interactions of the St\"uckelberg fields with gravity:
\beq \mathcal{L}'=\sqrt{-g}\,G^{\m\n}\left(\mathcal{\hat X}_{\m\n}-\tfrac{\a}{2}\mathcal{Y}_{\m\n}\right),\eeq{St5}
where the tensors $\mathcal{\hat X}_{\m\n}$ and $\mathcal{Y}_{\m\n}$ are given by
\bea &\mathcal{\hat X}_{\m\n}=-\tfrac{1}{2}\left(B_\m B_\n-\tfrac{1}{2}g_{\m\n}B^2\right)+\tfrac{\b}{\L^2}\left[2\nb_{[\m}B_{\r]}
\left(\nb_\n B^\r-\nb^\r B_\n\right)-g_{\m\n}\left(\nb_{[\r}B_{\s]}\right)^2\right],&\label{St4.95}\\
&\mathcal{Y}_{\m\n}=B_\m B_\n-\tfrac{2}{m}B_{(\m}\de_{\n)}\phi+\tfrac{1}{m^2}\de_\m\phi\de_\n\phi\,.&\eea{St5.0}

Let us now consider gravitational fluctuations around flat space:
\beq g_{\mu\nu}=\eta_{\mu\nu}+\tfrac{2}{M_\text{P}}h_{\mu\nu},\eeq{g1}
where the graviton field $h_{\mu\nu}$ is canonically normalized with mass dimension one. This helps assigning canonical dimensions to various operators in the gauge-fixed Lagrangian~(\ref{St4}).
The Einstein-Hilbert part of the Lagrangian reduces to a free quadratic part:
\beq\mathcal{L}_{\text{EH}}^{(\text{free})}=\tfrac{1}{2}h_{\m\n}\mathcal G^{\m\n},\qquad \mathcal{G}_{\m\n}\equiv\bbox h_{\m\n}-2\de_{(\m}\de^\r h_{\n)\r}+\de_\m\de_\n h^\r_\r
-\h_{\m\n}\left(\bbox h^\r_\r-\de^\r\de^\s h_{\r\s}\right),\eeq{E-H}
plus Planck-suppressed graviton self interactions given in Eq.~(\ref{St15}) in the Appendix. Expanding $\sqrt{-g}g^{\m\n}$ as in Eq.~(\ref{St312a}),  up to a total derivative
term one rewrites $\mathcal{L}_0$ as:
\beq \mathcal{L}_0\,=\,\tfrac{1}{2}\phi\left(\de^2-m^2\right)\phi+\tfrac{1}{M_P}\left[h^{\m\n}\de_\m\phi\de_\n\phi-\tfrac{1}{2}h^\m_\m\left(\de_\r\phi\de^\r\phi
+m^2\phi^2\right)\right]+\mathcal{O}\left(h^2\right).\eeq{St32.5}
In order to rewrite $\mathcal{L}_1$, we would need the covariant derivative expansion~(\ref{St101}) and the Riemann tensor expansion:
$R^{\m\r}{}_{\n\s}=-\tfrac{4}{M_P}\de^{[\m}\de_{[\n}h^{\r]}{}_{\s]}+\mathcal{O}(h^2)$. Thereby we have:
\bea \mathcal{L}_1=\tfrac{1}{2}B_\m\left(\de^2-m^2\right)B^\m+\tfrac{1}{M_P}h^{\m\n}\left(\de_\m B_\r\de_\n B^\r+\de_\r B_\m\de^\r B_\n
-\tfrac{1}{2}\h_{\m\n}(\de_\r B_\s)^2+m^2B_\m B_\n\right)\nonumber\\
+\tfrac{1}{M_P}\left(2\de^{(\mu}h^{\n)\r}-\de^\r h^{\m\n}\right)B_\r\de_\m B_\n-\tfrac{m^2}{2M_P}h^\m_\m B^2
+\tfrac{4\b}{\L^2M_p}\de^\m\de^\n h^{\r\s}\de_{[\m} B_{\r]}\de_{[\n} B_{\s]}+\mathcal{O}\left(h^2\right).~\eea{St31.5}

On the other hand, the cubic cross-interaction terms arising from Eq.~(\ref{St5}) are proportional to the linearized graviton equations of motion:
\beq \mathcal{L}'=-\tfrac{1}{M_P}\mathcal{G}^{\m\n}\left(\mathcal{X}_{\m\n}-\tfrac{\a}{2}\mathcal{Y}_{\m\n}\right)+\mathcal{O}\left(h^2\right),\eeq{Ricci-exp}
where $\mathcal{X}_{\m\n}$ is the flat-space counterpart of $\mathcal{\hat X}_{\m\n}$ given in Eq.~(\ref{St5.0}). These cubic terms can be eliminated by the following field redefinition:
\beq h_{\m\n}\rightarrow h_{\m\n}+\tfrac{1}{M_P}\left(\mathcal{X}_{\m\n}-\tfrac{\a}{2}\mathcal{Y}_{\m\n}\right).\eeq{St10}
To be explicit, after the field redefinition~(\ref{St10}), one gets:
\beq \mathcal{L}_{\text{EH}}^{(\text{free})}+\mathcal{L}'~\rightarrow~\mathcal{L}_{\text{EH}}^{(\text{free})}-\tfrac{1}{2M_P^2}\left(\mathcal{X}_{\m\n}-\tfrac{\a}{2}\mathcal{Y}_{\m\n}\right)
\mathcal{K}^{\m\n\r\s}\left(\mathcal{X}_{\r\s}-\tfrac{\a}{2}\mathcal{Y}_{\r\s}\right)+\mathcal{O}(h^2),\eeq{St10.1}
where $\mathcal{K}^{\m\n\r\s}$ is the quadratic differential operator in the graviton kinetic term, i.e.,
\beq \mathcal K^{\m\n\r\s}=\left(\h^{\m\n,\r\s}-\h^{\m\n}\h^{\r\s}\right)\de^2+\h^{\m\n}\de^\r\de^\s+\h^{\r\s}
\de^\m\de^\n-\h^{\m(\r}\de^{\s)}\de^\n-\h^{\n(\r}\de^{\s)}\de^\m\,,\eeq{g10}
with $\h^{\m\n,\r\s}\equiv\tfrac{1}{2}\left(\h^{\m\r}\h^{\n\s}+\h^{\m\s}\h^{\n\r}\right)$.
Note that Eq.~(\ref{St10.1}) encodes quartic interaction terms among the St\"uckelberg modes\footnote{Because the operator~(\ref{g10}) has zero modes of the form
$\de_{(\m}\l_{\n)}$, where $\l_\m$ is an arbitrary space-time function, the quartic interactions have different equivalent forms that differ by zero-mode contributions.}
through operators up to dimension 10.

\subsection{Cutoff Estimation}

In order for the effective field theory to make sense in the first place, it is  essential to have $m\ll M_\text{P}$. The high energy regime we are interested in
is characterized by the center-of-mass energy $m\ll\sqrt s\ll M_\text P$. Let us define the following mass scale:
\beq \L_3\equiv\sqrt[3]{m^2M_{\text P}}\,,\qquad \L_2\equiv\sqrt{mM_{\text P}}\,,\qquad m\ll\L_3\ll\L_2\ll M_P.\eeq{scales}

Now that all fields have canonical dimension one and the propagators are nonsingular in the limit of vanishing Proca mass, the interaction terms are clearly
non-renormalizable. Different nontrivial higher-dimensional operators may be suppressed by different mass scales. It is the lowest of these scales that defines
the ultraviolet cutoff of the theory. In the following we consider different cases of interest depending on the model.

$\bullet$~\underbar{Case I, $\a\neq 0$, No Counter Terms Added}: In this case, if we forgo the quadrupole term, the lowest suppression scale
turns out to be $\L_3$. In other words, $\L_3$-suppressed irrelevant operators become the most dangerous in the high-energy limit. Note that when the field
redefinition~(\ref{St10}) is implemented in either the graviton self coupling~(\ref{St15}) or $\mathcal{L}_0$ or $\mathcal{L}_1$, the resulting higher dimensional
operators are suppressed only by a scale $\L_2$ or higher. The same happens with the various $\mathcal{O}(h^2)$ terms we did not spell out. If we now include the
quadrupole term, it is required that $\L\gtrsim\L_3$. Let us take the decoupling limit:
\beq m\rightarrow 0,\qquad M_P\rightarrow \infty, \qquad \Lambda_3=\mbox{constant}.\eeq{limit00}
Then the theory does not become free, but reduces to the following simple Lagrangian:
\beq \mathcal{L}_{\text{g.f.}}~\rightarrow~\tfrac{1}{2}B_\m\bbox B^\m+\tfrac{1}{2}\phi\bbox\phi+\tfrac{1}{2}h_{\m\n}\mathcal G^{\m\n}
-\tfrac{\a^2}{8\L_3^6}\,\de_\m\phi\,\de_\n\phi\,\mathcal K^{\m\n\a\b}\de_\a\phi\,\de_\b\phi\,.\eeq{ST400}
Without counter terms, the cutoff $\L$ of the effective field theory is therefore given by:
\beq \L\sim\frac{\L_3}{\sqrt[3]{|\a|}}\,,\qquad 0<|\a|\lesssim\mathcal{O}(1).\eeq{cutoff10}
It is reassuring to note that the dimension-10 operators appearing in Lagrangian~(\ref{ST400}) are nontrivial quartic interactions, i.e.,
they cannot be removed by field redefinitions.

$\bullet$~\underbar{Case II, $\a\neq0$, Counter Terms Added}: It is possible to push the cutoff scale beyond~(\ref{cutoff10}) by adding suitable local counter terms.
We would like to cancel the $\L_3$-suppressed quartic scalar interaction term in Lagrangian~(\ref{St10.1}). The feat can be achieved by a counter
term of the following form (in the unitary gauge):
\beq \mathcal{L}_{\text{c.t.}}=\tfrac{1}{8}\k\a^2M_P^{-2}\sqrt{-g}\,A_\m A_\n\hat{\mathcal{K}}^{\m\n\r\s}A_\r A_\s,\eeq{CT0}
where $\hat{\mathcal{K}}^{\m\n\a\b}$ is a covariant counterpart of~(\ref{g10}). After the substitution~(\ref{St1}) is made in the above counter term, the desired
cancelations happen if we set
\beq\k=1.\eeq{kappa}
This actually eliminates all the quartic interaction terms originating from~(\ref{St10.1}) that contain at least three scalars. Then, forgoing the quadrupole term,
one is left with higher-dimensional operators suppressed only by a scale $\L_2$ or higher. In particular, the analogue of Eq.~(\ref{St10.1}) that spells out the
4-St\"uckelberg interaction terms reads:
\beq \mathcal{L}_{\text{EH}}^{(\text{free})}+\mathcal{L}'+\mathcal{L}_{\text{c.t.}}~\rightarrow~\mathcal{L}_{\text{EH}}^{(\text{free})}-\tfrac{1}{2M_P^2}\mathcal{X}_{\m\n}
\mathcal{K}^{\m\n\r\s}\left(\mathcal{X}_{\r\s}-\a\mathcal{Y}_{\r\s}\right)+\mathcal{O}(h^2).\eeq{St20}
Inclusion of the quadrupole term will then require $\L\gtrsim\L_2$. Because the $\L_2$-suppressed irrelevant operators are the most dangerous at high energies,
we take the decoupling limit:
\beq m\rightarrow 0,\qquad M_P\rightarrow \infty, \qquad \Lambda_2=\mbox{constant}.\eeq{limit100}
In this limit, our Lagrangian reduces to the following form:
\beq \mathcal{L}_{\text{g.f.}}+\mathcal{L}_{\text{c.t.}}\rightarrow\tfrac{1}{2}B_\m\bbox B^\m+\tfrac{1}{2}\phi\bbox\phi+\tfrac{1}{2}h_{\m\n}\mathcal G^{\m\n}
+\sum_{n=1}^\infty\left(\tfrac{\a}{\L_2^4}\right)^n\left(\mathcal{O}_{4n+4}+\b\L^{\!-2}\mathcal{O}_{4n+6}\right),\eeq{ST4}
where $\mathcal{O}_d$ denotes an operator of mass dimension $d$. The infinite series of higher-dimensional operators originates from the non-linearity in the graviton fluctuations.
At the interacting level, $(n+2)$-point couplings may contain $n$ canonically normalized gravitons, each of which appears with a factor $M_P^{-1}$. Then, the field redefinition~(\ref{St10})
of the graviton may bring an additional factor of $\left(m^2M_P/\a\right)^{-n}$. As a result, in the limit~(\ref{limit100}) one is left with an infinite series of irrelevant operators
in steps of mass dimension four.

Because $|\b|\sim\mathcal{O}(1)$, it is clear from Lagrangian~(\ref{ST4}) that the addition of the local counter term~(\ref{CT0})--(\ref{kappa}) has pushed
the effective field theory cutoff to a higher scale of
\beq \Lambda\sim\frac{\Lambda_2}{\sqrt[4]{|\a|}}\,,\qquad 0<|\a|\lesssim\mathcal{O}(1).\eeq{cutoff1000}

We will now prove that $\L$ given in Eq.~(\ref{cutoff1000}) is the upper bound of the effective field theory cutoff for $\a\neq0$. It suffices to show that
among the irrelevant operators in the Lagrangian~(\ref{ST4}) there is at least one that can neither be field redefined away nor can be canceled up to a total
derivative by local counter terms without worsening the ultraviolet behavior. With this end in view, let us single out the dimension-8 quartic
interaction terms of 2 vectors and 2 scalars. The origin of such terms is twofold: the graviton field redefinition~(\ref{St10}) acting on the
vector-graviton-vector cubic couplings in Eq.~(\ref{St31.5}), and the 4-St\"uckelberg interaction terms in Eq.~(\ref{St20}). These terms are captured, up to
total derivatives, by the following dimension-8 operator:
\beq \mathcal{O}_8=-\tfrac{1}{2}\left[\de^2(B_\m B_\n)\de^\m\phi\de^\n\phi+\!\de^2(B_\m\de^\m\phi)B_\n\de^\n\phi+\!\tfrac{1}{4}\de^2(B_\m\phi)\de^2(B^\m\phi)
\!-\!\tfrac{1}{8}\de^2B^2\de^2\phi^2\right]\!.\eeq{2v2s}
In deriving the above result we have dropped terms containing $\de^2B_\m$ or $\de^2\phi$ since we are interested in on-shell scattering amplitudes.
It is clear that the quartic interactions~(\ref{2v2s}) are nontrivial, i.e. they cannot all be eliminated by field redefinitions modulo total derivatives.
On the other hand, cancellation of the operator~(\ref{2v2s}) by local counter terms necessarily introduces new dimension-9 and dimension-10 operators which
make the ultraviolet behavior worse. This can be seen by noting that, in unitary gauge, candidate counter terms have the schematic form $M_P^{-2}\de^2A^4$.
After the substitution~(\ref{St1}), one will not only obtain the desired $\L_2^{-4}\de^4B^2\phi^2$ terms but also nontrivial terms like $\L_2^{-2}\L_3^{-3}\de^5B\phi^3$
and $\L_3^{-6}\de^6\phi^4$, which are suppressed by scales much smaller than $\L_2$. This completes the proof.

$\bullet$~\underbar{Case III, $\a=0$}: The situation changes drastically when $\a=0$. In this case, the original Lagrangian~(\ref{L0}) acquires a $U(1)$ gauge invariance
in the massless limit. This means that all the $1/m$ dependencies in the gauge-fixed Lagrangian~(\ref{St4}) must disappear. An inspection of the various terms in the
Lagrangian confirms this. More precisely, the tensor $\mathcal{Y}_{\m\n}$ given in Eq.~(\ref{St5.0}) never appears since it is accompanied by a coefficient of $-\a/2$.
Subsequently, the field redefinition~(\ref{St10}) may produce Planck-suppressed operators only. Therefore, the cutoff in this case is nothing but the Planck scale itself:
\beq \Lambda\sim M_P\,,\qquad \a=0.\eeq{cutoff200}
This is also the \emph{model-independent} upper bound on the ultraviolet cutoff~\cite{Rahman:2009zz,Rahman:2011ik,Kurt} of the Einstein-Proca system, as already mentioned.

\section{Shock-Wave Analysis}\label{sec:shockwave}

In this section we show, among other things, that the generalized Einstein-Proca model~(\ref{L0}) admits pp-wave solutions~\cite{Bondi:1958aj,Dray:1984ha}. Note that
similar solutions were found in~\cite{Gurses:1978zs} for generalized Einstein-Maxwell theories~\cite{Horndeski:1976gi}. We consider linear fluctuations of
the Proca field on this background in the probe approximation to demonstrate that, upon crossing the pp-wave, the fluctuations may undergo negative time delays
unless the parameter space of the theory is appropriately constrained. Such arguments have already been used for constraining higher-derivative gravity~\cite{Camanho:2014apa}
and massive gravity theories~\cite{Camanho:2016opx,Hinterbichler:2017qyt}.

\subsection{pp-Wave Solution}

Let us introduce the light-cone coordinate system $\left(u,v,\vec{x}\right)$, where
$u=t-x_3$, $v=t+x_3$, and $\vec{x}=\left(x_1,x_2\right)$. Then, a generic pp-wave spacetime has the following metric:
\beq \mathrm{d}s^2=-\mathrm{d}u\mathrm{d}v+\mathcal{F}(u,\vec{x})\mathrm{d}u^2+\mathrm{d}\vec{x}^2\,.\eeq{metric0}
This geometry enjoys the null Killing vector $\partial_v$. One can introduce a covariantly constant null vector $l_\mu=\delta_{\mu u}$
to write this metric in the Kerr-Schild form:
\beq \bar g_{\mu\nu}=\eta_{\mu\nu}+\mathcal{F}(u,\vec{x})\,l_\mu l_\nu\,.\eeq{metric1}
The inverse and the Christoffel symbols corresponding to the metric~(\ref{metric1}) are given by
\beq \bar g^{\m\n}=\h^{\m\n}-\mathcal{F}l^\m l^\n,\qquad \bar\G^\l_{\m\n}=l^\l l_{(\m}\de_{\n)}\mathcal{F}-\tfrac{1}{2}l_\m l_\n\de^\l \mathcal{F},\eeq{christ}
which in turn yield the following curvature quantities:
\beq \bar R^\r{}_{\s\m\n}=l_\s l_{[\m}\de_{\n]}\de^\r\mathcal{F}-l^\r l_{[\m}\de_{\n]}\de_\s\mathcal{F},\qquad
\bar R_{\m\n}=-\tfrac{1}{2}l_{\m}l_{\n}\de^2 \mathcal{F},\qquad \bar R=0.\eeq{curvature}

To see if the generalized Einstein-Proca system~(\ref{L0}) admits pp-wave solutions, let us first write down the resulting equations of motion. The Einstein
equations read:
\beq G_{\mu\nu}=M_P^{-2}T_{\m\n}\equiv M_P^{-2}\left[T_{\m\n}^{(0)}+\a T_{\m\n}^{(\a)}+\b T_{\m\n}^{(\b)}\right],\eeq{eom1}
where the various parts comprising the stress-energy tensor are given by:
\bea &T_{\m\n}^{(0)}=F_{\m\r}F_\n{}^\r-\tfrac{1}{4}g_{\m\n}F_{\r\s}^2+m^2\left(A_\m A_\n-\tfrac{1}{2}g_{\m\n}A^2\right)\!,&\nonumber\\
&T_{\m\n}^{(\a)}=\tfrac{1}{2}\left[\d^\r_{\m}\d^\s_{\n}(\nb^2\!-\!R)\!+\!g_{\m\n}(\nb^{\r}\nb^{\s}\!\!-\!G^{\r\s})\!-\!2\d_{(\m}^\r\nb_{\n)}\nb^{\s}\!+\!g^{\r\s}R_{\m\n}\right]
\!\left(A_\r A_\s\!-\!\tfrac{1}{2}g_{\r\s}A^2\right)\!,~~~~&\label{ProcaT}\\
&T_{\m\n}^{(\b)}=-\L^{\!-2}\left(L_{\m\r\n\s}F^{\r\l}F^{\s}{}_{\l}+\nb_\r\tilde{F}_{\s\m}\nb^\s\tilde{F}^\r{}_{\n}\right)\!,&\nonumber\eea{yyyyyy}
with $\tilde{F}^{\m\n}=\tfrac{1}{2}\ve^{\m\n\r\s}F_{\r\s}$ being the dual Faraday tensor. The Proca equations of motion are:
\beq \left(g^{\m\s}g^{\n\r}+\b\L^{\!-2}L^{\m\n\r\s}\right)\nb_\n F_{\r\s}-\left(m^2g^{\m\n}+\a\,G^{\m\n}\right)A_\n=0.\eeq{eom2}
The system of equations~(\ref{eom1})--(\ref{eom2}) admits the solution:
\beq g_{\m\n}=\bar{g}_{\m\n}=\h_{\m\n}+\mathcal{F}(u,\vec{x})\,l_\m l_\n\,,\qquad A_\m=\bar{A}_\m=M_P\mathcal{H}(u,\vec{x})\,l_\m\,,\eeq{sw1}
provided that the functions $\mathcal{F}$ and $\mathcal{H}$ satisfy the following equations:
\beq \de^2\tilde{\mathcal F}=0\,,\qquad \left(\de^2-m^2\right)\mathcal{H}=0\,,\eeq{Poisson}
where the function $\tilde{\mathcal F}$ has been defined as
\beq\tilde{\mathcal F}\equiv\mathcal{F}+\left(1+\a+2\b\L^{\!-2}m^2-\tfrac{1}{2}\b\L^{\!-2}\de^2\right)\mathcal{H}^2\,.\eeq{defined-F}

Because $\tilde{\mathcal{F}}$ and $\mathcal{H}$ are independent of $v$, the d'Alembertian operator in Eq.~(\ref{Poisson}) can be replaced by
the Laplace operator on the transverse plane. Assuming rotational symmetry on the transverse plane, the equations have the following solutions
at $\vec x \neq 0$:
\beq \tilde{\mathcal{F}}=-\mathcal{A}(u)\ln(\L|\vec{x}|)\,,\qquad \mathcal{H}=\mathcal{B}(u)K_0(m|\vec{x}|)\,,\eeq{sol1}
where $\mathcal{A}(u)$ and $\mathcal{B}(u)$ are \emph{arbitrary} functions of $u$, and $K_0$ is the zeroth-order modified Bessel function of the second kind.
Note that the singularity of the solutions~(\ref{sol1}) at $\vec x=0$ can be attributed to delta function-type sources. To be explicit, one could consider
the following source current on the right-hand side of Eq.~(\ref{eom2}):
\beq j^\m=-2\pi M_P \mathcal{B}(u)\delta^2(\vec x)\,l^\m\,.\eeq{source-current}
Similarly, the stress-energy tensor in Eq.~(\ref{eom1}) ought to include a singular piece:
\beq M_P^{-2}T_{\mu\nu}=\pi l_\m l_\n\mathcal{A}(u)\delta^2(\vec x)\!+\!m^2l_\m l_\n \mathcal{B}^2(u)\!\left[K^2_0(m|\vec{x}|)\!+\!K^2_1(m|\vec{x}|)\right]\!
\left[1\!+\!\a-\!\b\,\mathfrak{F}(m|\vec{x}|)\right],
\eeq{stress-energy}
where we have defined the function:
\beq \mathfrak{F}(m|\vec{x}|)\equiv\tfrac{1}{2}\L^{\!-2}\left[K^2_0(m|\vec{x}|)+K^2_1(m|\vec{x}|)\right]^{-1}\de^2K^2_1(m|\vec{x}|)-m^2\L^{\!-2}.\eeq{whatisthis}

Let us now find the consequences of imposing null-energy condition on the stress-energy tensor, i.e., $T_{\m\n}n^\m n^\n\geq0$ for any null vector $n^\m$. Because
$\mathcal A(u)$ and $\mathcal B(u)$ are a priori arbitrary functions, it immediately follows that:
\beq \mathcal{A}(u)\geq0,\qquad 1+\a-\b\,\mathfrak{F}(m|\vec{x}|)\geq0.\eeq{cond101}
Further conditions arise from noting that $\mathfrak{F}(m|\vec{x}|)$ is a positive-definite function. In the effective field theory context, however, it only makes
sense to talk about distances no smaller than $1/\L$. Now, starting from the value of $\mathfrak{F}(m/\L)$ equal to 2, the function decreases monotonically as $|\vec{x}|$
increases, and approaches zero as $|\vec x|\rightarrow\infty$.
Then, within the effective field theory, positive semi-definiteness of $\mathfrak{F}(m|\vec{x}|)$ is tantamount to:
\beq 1+\a-2a\b\geq0,\quad \text{for}~a\in(0,1].\eeq{cond102}
As already noted in~\cite{{Dubovsky:2005xd,Nicolis:2009qm,Rubakov:2014jja}}, the violation of null-energy condition may give rise to superluminal propagation.
In the next two subsections we will consider probe fluctuations on the background~(\ref{sw1}) and compute the time delays they suffer upon crossing
the pp-wave.

\subsection{Linear Fluctuations}

On the background~(\ref{sw1}) let us now consider, in the probe limit, linear Proca fluctuations:
\beq w_\m=A_\m-\bar A_\m.\eeq{probe}
Derived easily from Eq.~(\ref{eom2}), the equations of motion of these fluctuations read:
\beq \left(\bar{\nb}^2-m^2\right)w_\m-\bar\nb_\m\left(\bar\nb\!\cdot\!w\right)-(\a+1)\bar G_{\m\n}w^\n+2\b\L^{\!-2}\bar L_{\m\n\r\s}\bar{\nb}^\n\bar{\nb}^\r w^\s=0,\eeq{eom211}
where barred quantities are constructed from and index contractions are performed by the background metric $\bar{g}_{\m\n}$ and its inverse.
A divergence of the equations of motion gives:
\beq \bar\nb\!\cdot\!w=\tfrac{1}{2}\a\,m^{\!-2}\de^2\mathcal{F}\,l\!\cdot\!\de\,l\!\cdot\!w,\eeq{sw2}
thanks to the transversality properties of $G_{\m\n}$ and $L_{\m\n\r\s}$. The divergence constraint~(\ref{sw2}) renders one component of the vector field $w_\m$
non-dynamical, leaving us with 3 dynamical degrees of freedom as expected. More specifically, one can rewrite Eq.~(\ref{sw2}) as:
\beq \de_v w_u=\tfrac{1}{2}\de_iw_i-\de_u w_v-\left(2\mathcal F+\a m^{\!-2}\de^2\mathcal F\right)\de_v w_v.\eeq{const111}
Because its $v$-derivative is completely determined by the other components, $w_u$ is clearly non-dynamical if one chooses $v$ as the light-cone time, as we will do.

The true dynamics of the system is found by substituting the constraint~(\ref{sw2}) in the equations of motion~(\ref{eom211}). The result is:
\beq \left(\de^2-m^2\right)w_\m\equiv\left(\bar{g}^{\r\s}\de_\r\de_\s-m^2\right)w_\m=\d\mathcal{R}_\m+l_\m\d\mathcal{R},\eeq{sw3}
where the quantities $\d\mathcal{R}_\m$ and $\d\mathcal{R}$ depend on the curvature through $\mathcal F$. Explicitly,
\begin{equation}\label{sw4}
\begin{split}
&\d\mathcal{R}_\m=4\de_\m\mathcal F\,\de_v w_v+\tfrac{2\a}{m^2}\,\de_\m\!\left(\de^2\mathcal F\de_v w_v\right)
+\tfrac{8\b}{\L^2}\left(\de_\m\de_\r\mathcal F-\bar g_{\m\r}\de^2\mathcal F\right)\!\left(\de_v^2 w^\r-\de^\r\de_v w_v\right),\\
&\d\mathcal{R}=2\de^\r\mathcal F\left(\de_v w_\r-\de_\r w_v\right)+\a\de^2\mathcal F\,w_v
+\tfrac{4\b}{\L^2}\left(\de^\r\de^\s\mathcal F-\bar g^{\r\s}\de^2\mathcal F\right)\!\left(\de_v\de_\r w_\s-\de_\r\de_\s w_v\right).
\end{split}
\end{equation}
When we take the $u$-component of the equations of motion~(\ref{sw3}), we get
\beq \left(\de_i^2-m^2\right)w_u=4\left(\de_u+\mathcal F\de_v\right)\de_v w_u+\d\mathcal{R}_u+\d\mathcal{R}.\eeq{u-comp}
It is easy to see that in the right-hand side of the above equation $w_u$ appears always through its $v$-derivative, and so it
can be completely eliminated by virtue of the constraint~(\ref{const111}). This simply reconfirms the status of $w_u$ as a non-dynamical variable.

The dynamical equations correspond to the $\m\!=\!v$ and $\m\!=\!i$ components of the equations of motion~(\ref{sw3}). They take the form:
\begin{equation}\label{v-comp}
\begin{split}
\left(\h^{\r\s}\de_\r\de_\s-m^2\right)w_v&=Y\de^2_v w_v+Y_i\de_v w_i\,,\\
\left(\h^{\r\s}\de_\r\de_\s-m^2\right)w_i&=Z_i\de_v w_v+Z_{ij}\de_v^2w_j\,,
\end{split}\end{equation}
where we have introduced the following functions and operators:
\bea
&Y=4\!\left(\mathcal F+\tfrac{\a}{2m^2}\de^2\mathcal F\right),\qquad Y_i=0,&\nonumber\\
&Z_i=4\de_i\!\left(\mathcal F+\tfrac{\a}{2m^2}\de^2\mathcal F\right)+4\!\left(\tfrac{\a}{2m^2}+\tfrac{2\b}{\L^2}\right)\!\de^2\mathcal F\de_i
-\tfrac{8\b}{\L^2}\,\de_i\de_j\mathcal F\de_j,&\label{operators22}\\
&Z_{ij}=4\!\left(\mathcal F-\tfrac{2\b}{\L^2}\de^2\mathcal F\right)\!\d_{ij}+\tfrac{8\b}{\L^2}\,\de_i\de_j\mathcal F.
\nonumber\eea{}
In what follows, we will assume that the fluctuations do not propagate through $\vec x=0$, so that the background equations~(\ref{Poisson}) can be used.

\subsection{Shapiro Time Delay}

The Proca-fluctuation modes will experience Shapiro time delay~\cite{Shapiro:1964uw} as they cross the pp-wave. Before computing this quantity, let us
specify the $u$-profiles of our background solution~(\ref{sol1}). We will choose the following ``sandwich wave''~\cite{Bondi:1958aj} profile:
\beq \left(
       \begin{array}{c}
         \mathcal{A}(u) \\
         \mathcal{B}(u) \\
       \end{array}
     \right)
=\left(
       \begin{array}{c}
         \mathcal{A}_0 \\
         \mathcal{B}_0 \\
       \end{array}
     \right)\left[1-\theta\!\left(u^2-\l^2\right)\right]\exp\left[-\frac{\lambda^2 u^2}{(u^2-\lambda^2)^2}\right]\,,\eeq{shock-F}
where numerical constants $\mathcal{A}_0$ and $\mathcal{B}_0$ define the amplitude, and the length scale $\lambda$ defines the width of the smeared wave.
We have chosen $\mathcal A(u)$ and $\mathcal B(u)$ to be smooth functions, $\mathcal A(u), \mathcal B(u)\in \mathrm{C}^{\infty}(\mathbb{R})$, with a compact
support $[-\lambda,\lambda]$. Note that the sandwich wave moves at the speed of light in the $v$-direction.
For future convenience, we introduce yet another pair of numerical constants $\bar{\mathcal{A}}_0$ and $\bar{\mathcal{B}}_0$, defined as:
\beq \bar{\mathcal{A}}_0\equiv\lambda^{-1}\!\int_{-\lambda}^{+\lambda}\mathrm{d}u\,\mathcal A(u)\approx 1.07\mathcal{A}_0,\qquad
\bar{\mathcal{B}}_0^{2}\equiv\lambda^{-1}\!\int_{-\lambda}^{+\lambda}\mathrm{d}u\,\mathcal B^2(u)\approx 1.13\,\mathcal{B}_0^2.\eeq{normalization}

Let us write the general solutions of the equations of motion~(\ref{sw3}) and the constraint~(\ref{sw2}) as superpositions of eigensolutions of the form:
\beq w_\m(u,v,\vec{x})=\tilde{w}_\m(u)\,e^{i\left(pv+\vec{q}\cdot\vec{x}\right)}\,,\eeq{Fourier}
where $p$ and $\vec q$ are the momenta in the $u$- and the transverse directions respectively. Because the probe experiences a radial impulse in
the transverse plane during the course of the sandwich wave, $u\in[-\lambda,\lambda]$, the transverse momenta will be $u$-dependent: $\vec q=\vec q(u)$.
We denote the incoming and outgoing transverse momenta by $\vec q_{_-}$ and $\vec q_{_+}$ respectively. Let the impact parameter vector in the transverse
plane at $u=-\lambda$~be~$\vec b=|\vec{b}|\,\vec e=b\,\vec e$. For the incoming transverse momenta we make the choice: $\vec q_{_-}=q_{_-}\vec e$~\,with~\,$q_{_-}\!>0$.

Given that there is a huge separation between the effective field theory cutoff $\Lambda$ and the Proca mass $m$, it is possible to accommodate
the following parametric relations:
\beq \Lambda\gtrsim\frac{1}{\lambda}\gg p\gg q_{_-}\gg\frac{1}{b}\gg m\,.\eeq{parameter-regime}
The reasons we are interested in the regime~(\ref{parameter-regime}) are the following.
The condition $q_{_-}b\gg1$ takes into account the requirement that the probe is far away from $\vec x=0$. The particle is chosen to be ultra-relativistic,
$p\gg q_{_-}\gg m$, for the sake of simplicity. All its momenta are however much smaller than $\Lambda$. The sandwich wave, on the other hand,
is taken to be thinner than all the characteristic length scales of the probe: $\lambda p\ll1$, but thick enough to be ``seen'' by the effective theory:
$\lambda\Lambda\gtrsim1$. The small impact parameter, $mb\ll1$,  is meant for amplifying the effects of the sandwich wave on the probe.

The change in transverse position while the probe particle is passing through the sandwich wave is small: $|\vec x-\vec b|\lesssim\lambda$,
which we will neglect. The radial impulse deflects the particle but keeps $\vec q(u)$ aligned with $\vec e$: $\vec q(u)=q(u)\vec e$. Note that $q(u)$ remains
positive and small compared to $p$. This can be seen by using the deflection formula~(A.36) of Ref.~\cite{Dray:1984ha}, which is a valid approximation because
the sandwich wave is thin. With energy $E\sim M_\textrm{P}^2\lambda$ of the sandwich wave and $\vec q_{_+}\equiv q_{_+}\vec e$, we can write
$(q_{_-}/p)-(q_{_+}/p)\sim\lambda/b$.
Given the separation of scales~(\ref{parameter-regime}), we conclude that $q_{_+}\!>0$~\,and~\,$q_{_+}\approx q_{_-}$. The same conclusion holds for $q(u)$
as it varies continuously. The unit transverse position vector, $\vec n\equiv\vec x/|\vec x|$, always coincides with $\vec{e}$ in our setup.

Let us redefine the dynamical modes as
\beq \Phi_1=\tilde{w}_v\,,\qquad \Phi_2=\d_{ij}e_i\tilde{w}_{j}\,,\qquad \Phi_3=\varepsilon_{ij}e_i\tilde{w}_{j}\,,\eeq{modes-defined}
where $i,j=1,2$ are indices in the transverse plane and $\varepsilon_{ij}$ is the Levi-Civita symbol. In terms of the new dynamical fields, collectively denoted as
$\{\Phi_I(u)\}$ with $I=1,2,3$, the equations of motion~(\ref{v-comp}) can be rewritten as:
\beq \left(\partial_u-ip\gamma\right)\Phi_I(u)=ip\left(\mathcal A(u)\mathcal{C}_{IJ}+\mathcal B^2(u)\mathcal{D}_{IJ}\right)\Phi_J(u),\eeq{DE-0}
where $\gamma\equiv\tfrac{1}{4}(q^2+m^2)/p^2$, and the $3\hspace{-0pt}\times\hspace{-0pt}3$ matrices $\mathcal{C}_{IJ}$ and $\mathcal{D}_{IJ}$
depend, apart from the Lagrangian parameters $\a$, $\b$ and $\L$, on the mass $m$ and momenta $p$, $q$ of the probe, the impact parameter $b$
and the width $\l$ of the sandwich wave. Their explicit forms appear in the Appendix through Eqs.~(\ref{C-matrix})--(\ref{D-matrix2}).
The a priori arbitrary functions $\mathcal{A}(u)$ and $\mathcal{B}(u)$ have been chosen as~(\ref{shock-F}), but the values of the constants
$\mathcal{A}_0$ and $\mathcal{B}_0$ are at our disposal. In particular, we can set either one of them to be zero and still have a
nontrivial background solution. It serves our purpose to consider the following two choices.

$\bullet$~\underbar{Choice I, $\bar{\mathcal{A}}_0=\pm1,\,\mathcal{B}_0=0$}: In this case it is easy to diagonalize the set of first-order coupled differential
equations~(\ref{DE-0}). Note that the eigenvalues of $\mathcal C$ are given by:
\beq c_1=\ln\!\left(\L b\right),\qquad c_2=\ln\!\left(\L b\right)-2\b\left(\L b\right)^{\!-2},\qquad c_3=\ln\!\left(\L b\right)+2\b\left(\L b\right)^{\!-2},\eeq{eigenvalues0}
which are independent of the momenta $p$ and $q$. The matrix $\mathcal U$ composed of the eigenvectors of $\mathcal C$ is $u$-dependent, but only as weakly
as $q(u)/p$. Then, in terms of the modes $\Psi_I\equiv\mathcal{U}^{-1}_{IJ}\Phi_J$, Eqs.~(\ref{DE-0}) are approximately diagonal, and hence
can be integrated to
\beq \Psi_I(+\l)\,\approx\,\Psi_I(-\l)\exp\left[ip\int_{-\l}^{+\l}\mathrm{d}u\,\left[\g+c_I\mathcal A(u)\right]\right]\,.\eeq{integrated}

The integral in Eq.~\eqref{integrated} is to be understood as the shift in the $v$-coordinate suffered by the $I$-th mode upon crossing the sandwich
wave~\cite{Camanho:2014apa}. To find the shift relative to massless propagation in flat space, let us write the relevant terms originating from $\gamma$:
\beq \Delta\gamma=\tfrac{1}{4}m^2/p^2+\tfrac{1}{4}(q^2-q_{_-}^2)/p^2\,.\eeq{gamma-break}
The first piece comes from non-zero Proca mass, whereas the second from non-zero curvature. Then, the $v$-shifts relative to flat-space massless
propagation can be written as:
\beq \D v_I~\equiv~\int_{-\l}^{+\l}\mathrm{d}u\,\left[\D\g+c_I\mathcal A(u)\right]~\approx~\left(\text{sgn}\mathcal{A}_0\right)c_I\l.\eeq{v-shift}

A positive shift corresponds to a time delay, whereas a negative $\D v$ to a time advance. Because $\L b\gg1$ and $|\b|\sim\mathcal{O}(1)$,
it is clear from Eq.~(\ref{eigenvalues0}) that all the $c_I$'s are large positive numbers. Negative time delays can be avoided by requiring that
$\text{sgn}\mathcal{A}_0=+1$. This requirement already follows from null-energy condition, which sets $\mathcal{A}(u)>0$.

$\bullet$~\underbar{Choice II, $\mathcal{A}_0=0,\,\bar{\mathcal{B}}_0=\pm1$}: Here we can follow the logical steps of the previous choice almost verbatim.
The $v$-shifts relative to flat-space massless propagation are:
\beq \D v_I~\equiv~\int_{-\l}^{+\l}\mathrm{d}u\,\left[\D\g+d_I\mathcal B^2(u)\right]~\approx~d_I\l,\eeq{v-shift1}
where $d_I$ are the eigenvalues of the matrix $\mathcal{D}$, which should be positive semi-definite in order that negative time delays be absent.
It suffices to write down only leading-order terms of the $d_I$'s in the regime of interest~(\ref{parameter-regime}). For small impact parameters,
$\e\equiv mb\ll1$, the eigenvalues reduce to the simple form:
\beq d_1\approx\a\left(1+\a-2\b(\L b)^{\!-2}\right)\e^{-2}+(1+\a)^2\ln^2\e,\qquad d_2\approx d_3\approx(1+\a)\ln^2\e.\eeq{eigenvalues11}
We therefore require: $\a\left(1+\a-2\b(\L b)^{\!-2}\right)\geq0$ and $\left(1+\a\right)\geq0$. Given the requirement~(\ref{cond102}) from null-energy
condition, it follows that $\a$ must be constrained as: $\a\geq0$.

To summarize, requiring null-energy condition on the pp-wave background and the absence of negative time delays in high-energy scattering results in the
following set of necessary and sufficient conditions on the dimensionless parameters $\a$ and $\b$:
\beq \a\geq0,\qquad \b\leq\tfrac{1}{2}\left(1+\a\right).\eeq{alphacons}
This equation comprises one of our main results.

\section{Unitarity \& Analyticity Constraints}\label{sec:forward}

The scattering amplitudes of a low-energy effective field theory ought to satisfy certain inequalities in order for a standard local, unitary, analytic and
Lorentz-invariant ultraviolet completion to exist~\cite{Adams:2006sv}. Independent of the details of the ultraviolet physics, these conditions take a
simple form for \emph{crossing-symmetric} amplitudes in the \emph{forward limit}. Known as positivity constraints, they hinge on the compliance
of high-energy scattering amplitudes with the optical theorem, the Froissart bound~\cite{Froissart:1961ux,Martin:1962rt}, and the S-matrix analyticity
properties. Positivity constraints may put rigorous bounds on the parameter space of various effective field theories: Einstein gravity with higher-curvature
corrections~\cite{Bellazzini:2015cra}, ghost-free massive gravity~\cite{Cheung:2016yqr}, and pseudo-linear massive gravity~\cite{Bonifacio:2016wcb},
for example\footnote{It is possible to generalize the requirements to an infinite number of positivity bounds at and away from the forward scattering
limit~\cite{deRham:2017avq,deRham:2017zjm}, which too could constrain various effective field theories~\cite{deRham:2018qqo,deRham:2017imi}.
Constraints for particles with nonzero spin and general polarizations, albeit subtle because of nontrivial crossing relations, have also been
derived~\cite{deRham:2017zjm,Bellazzini:2016xrt}. See also~\cite{Bellazzini:2017fep} for bounds beyond positivity.}.

In this section, we will see that the positivity arguments~\cite{Adams:2006sv} are ineffective in constraining our Einstein-Proca effective theory~(\ref{L0}).
For the sake of completeness, we will also investigate the consequences of the counter term~(\ref{CT0}), keeping the parameter $\k$ arbitrary. We are interested
in on-shell 4-point scattering amplitudes of the Proca field that are simultaneously forward and invariant under crossing in the $t$-channel. Formally, $t$-channel
crossing symmetry is tantamount to the invariance under the particle-label swapping $1\leftrightarrow3$ or $2\leftrightarrow4$. For external Proca particles,
this is ensured if the exchanged particles have the same polarization relative to their momenta. Then, the forward limit corresponds to the following identification
of particles: $1\leftrightarrow3$ and $2\leftrightarrow4$.

The presence of dynamical gravity creates an obstacle since $t$-channel graviton exchange gives a singular contribution in the forward limit: $t\rightarrow0$
($s$, $t$ and $u$ are the Mandelstam variables). As a result, the Froissart bound is violated rendering the positivity arguments invalid. In this case an infrared
regulator $\m$ comes to rescue, as already noted in~\cite{Adams:2006sv,Cheung:2014ega,Bellazzini:2015cra}. The regulator$-$much smaller than any physical mass scale
in the theory$-$is introduced by sending $t\rightarrow t-\m^2$ in the amplitude. This alleviates the $t$-singularity in the forward limit as the
$t$-channel exchange now produces a large but finite contribution to the amplitude. Our forward amplitudes will therefore correspond to the following kinematic
regime of interest:
\beq \L\gg\sqrt s\gg\sqrt{-t}=\m\ll m.\eeq{alleviate}

We follow the procedure outlined in~\cite{Adams:2006sv,Cheung:2016yqr}, and assume a perturbative ultraviolet completion
of our theory, which allows for an $\hbar$-expansion and justifies the consideration of only tree-level diagrams. For technical details related to non-zero particle spin
and massless $t$-channel exchange we refer the readers to~\cite{Bellazzini:2015cra,Bellazzini:2016xrt}. We consider the 4-point scattering amplitudes of the process
$1+2\rightarrow 3+4$ involving on-shell Proca fields of \emph{definite helicity}: ${\cal M}_{\l_1\l_2;\,\l_3\l_4}(s,t)$, where the $\l_n$ is the polarization of the
$n$-th particle. We choose particles 1 and 2 as incoming, particles 3 and 4 as outgoing, and linear polarization basis $\l_n=0,1,2$, where $0$ denote the
longitudinal polarization, and $1$ and $2$ the transverse polarizations parallel and perpendicular to the scattering plane respectively.
Therefore, the infrared-regulated forward\footnote{Note that the derivation of positivity bounds does not require $t$ to be strictly
zero~\cite{Bellazzini:2015cra,Martin:1962rt}.} crossing-symmetric amplitudes of definite helicity are:
\beq {\cal M}_{ij}(s)={\cal M}_{ij;\,ij}\left(s,t\!\rightarrow\!-\m^2\right),\qquad i,j=0,1,2, \eeq{disp0}
where the particle identifications $1\leftrightarrow3$ and $2\leftrightarrow4$ have been made. Then, we consider the following quantity:
\beq f_{ij}\equiv\frac{1}{2\pi i}\oint_{\G}ds\,\frac{{\cal M}_{ij}(s)}{(s-s_0)^{3}}\,,\eeq{disp1}
where $s_0$ is an arbitrary point within the real-line segment $(0,4m^2)$ on the complex $s$-plane, and the small contour $\G$ encircles the pole at $s=s_0$.
From analytic dispersion relations~\cite{Adams:2006sv}, it turns out that the $f_{ij}$'s can be computed at tree level solely within the effective
theory\cite{Cheung:2016yqr, Bellazzini:2017fep}, as the negative residue of the integrand at large $s$:
\beq f_{ij}=-\underset{s=\infty}{\text{Res}}\left[\frac{{\cal M}_{ij}(s)}{(s-s_0)^3}\right]_{\text{EFT}}.\eeq{disp2}

Given that the Froissart bound holds on account of unitarity and locality$-$thanks to the infrared regulator$-$one can deform the contour $\mathcal C$ to encircle the
multi-particle branch cuts starting at $s=0$ and $s=4m^2$ dropping the boundary contribution, which vanishes at infinity. The value thus obtained is related to the
total cross-section by virtue of the optical theorem and crossing symmetry. Because the total cross-section is positive, one finds that the $f_{ij}$'s must be strictly
positive~\cite{Adams:2006sv,Cheung:2016yqr}:
\beq f_{ij}>0,\qquad i,j=0,1,2.\eeq{f-positive}

To calculate the quantities $f_{ij}$ in our model from Eq.~(\ref{disp2}), we recourse to the Mathematica packages \emph{xAct`xTensor'}, \emph{FeynRules}~\cite{Alloul:2013bka}
and \emph{FeynCalc}~\cite{Shtabovenko:2016sxi}. The results are independent of the arbitrary mass scale $\sqrt{s_0}$, and given by:
\begin{subequations}\label{f-given}
\begin{align}
f_{00}&=(mM_P)^{\!-2}\left[\,\left(m/\m\right)^2+\a\,\right],\label{f-00}\\
f_{10}&=(mM_P)^{\!-2}\left[\,\left(m/\m\right)^2+\tfrac{1}{2}\a-\tfrac{1}{4}\a^2(1-\k)\,\right],\label{f-10}\\
f_{20}&=(mM_P)^{\!-2}\left[\,\left(m/\m\right)^2+\tfrac{1}{2}\a-\tfrac{1}{4}\a^2(1-\k)-\b\left(m/\L\right)^2\,\right],\label{f-20}\\
f_{11}&=(\L M_P)^{\!-2}\left[\,\left(\L/\m\right)^2+\tfrac{1}{2}\b^2\left(m/\L\right)^2-\b\,\right],\label{f-given1}\\
f_{22}&=(\L M_P)^{\!-2}\left[\,\left(\L/\m\right)^2+\tfrac{1}{2}\b^2\left(m/\L\right)^2-3\b\,\right],\label{f-given2}\\
f_{12}&=(\L M_P)^{\!-2}\left[\,\left(\L/\m\right)^2+\tfrac{1}{2}\b^2\left(m/\L\right)^2-\b(2-\a)\right]=f_{21}.\label{f-given3}
\end{align}
\end{subequations}

We see that the $t$-channel graviton exchange dominates all the crossing-symmetric forward amplitudes through large \emph{positive} contributions of
$\mathcal{O}((m/\m)^2)$ and $\mathcal{O}((\L/\m)^2)$. Clearly, the positivity constraints~(\ref{f-positive}) do not give rise to any useful bounds on the parameters
$\a$ and $\b$; their contributions are washed out by that of the $t$-channel graviton exchange, and there is no parameter regime where the latter is subdominant.

A couple of remarks are in order. First, one may instead take indefinite-helicity amplitudes into consideration. For such amplitudes as well, it is easy to see that the
graviton-exchange contributions render the positivity constraints ineffective. Second, while we have made a nontrivial assumption of a perturbative ultraviolet completion,
it is well known that loop corrections in quantum gravity are typically infrared divergent~\cite{Barvinsky:1985an}. One may wonder whether there could be
cancellations among the singular contributions in the full amplitudes, to which the positivity bounds apply after all. While it is a daunting task to compute the full
amplitudes, we have investigated the 1-loop contributions to the 4-point scattering amplitudes under consideration. It is not difficult to be convinced that the
singular terms arising from such quantum corrections are suppressed by small numbers, $(\L/M_P)^2$ or $(m/M_P)^2$ for example, when compared to their tree-level counterparts.
So, we cannot conclude anything concrete about the effective field theory parameters.

\section{Summary \& Conclusions}\label{sec:remarks}

In this article we have studied the gravitational properties of a massive spin-1 field in the context of a two-parameter family of Einstein-Proca Lagrangians, which (i) is at most
quadratic in the Proca field, (ii) admits no higher-derivative terms in the equations of motion, (iii) incorporates all possible Proca-graviton-Proca cubic couplings in flat space,
and (iv) allows for an ambiguity term that arises naturally from minimal coupling. We have estimated the model-(in)dependent upper bound on the ultraviolet cutoff of the effective
field theory under consideration. The cutoff scale $\L$ depends on the Proca mass $m$, the Planck mass $M_P$ and the ambiguity parameter $\a$.

Let us clarify some points regarding the various estimates~(\ref{cutoff10}),~(\ref{cutoff1000}) and~(\ref{cutoff200}) of the cutoff scale in order that the results fit nicely
with each other. As already noted, it is technically natural to have a small value of $\a$. Na\"ively, the scales~(\ref{cutoff10}) and~(\ref{cutoff1000}) blow up in the limit
$\a\!\rightarrow\!0$. However, as soon as $|\a|$ becomes $\mathcal{O}((m/M_P)^2)$, the separation of the scales~(\ref{cutoff10}) and~(\ref{cutoff1000}) from the Planck mass 
disappear. Therefore, it makes more sense to summarize our cutoff estimates as:
\beq \L~\sim~\left\{
             \begin{array}{ll}
               \frac{\L_3}{\sqrt[3]{|\a|}}, & \hbox{$\mathcal{O}((m/M_P)^2)\lesssim|\a|\lesssim\mathcal{O}(1)$, without counter terms,} \\
               \frac{\L_2}{\sqrt[4]{|\a|}}, & \hbox{$\mathcal{O}((m/M_P)^2)\lesssim|\a|\lesssim\mathcal{O}(1)$, with counter terms,} \\
               M_P, & \hbox{$|\a|\lesssim\mathcal{O}((m/M_P)^2)$.}
             \end{array}
           \right.
\eeq{casess}
We emphasize that the model-independent upper bound on the ultraviolet cutoff is $M_P$. Of course in a given consistent Proca-Einstein theory the cutoff scale can
actually be much lower than this, e.g., Abelian Higgs model coupled to gravity with the Higgs field integrated out, in which case it is the Higgs mass that defines
the cutoff scale.

We have found pp-wave solutions of the Einstein-Proca model~(\ref{L0}). When subject to null-energy conditions, these background solutions give rise to nontrivial
constraints~(\ref{cond102}) on the parameters $\a$ and $\b$ of the effective theory. We have additionally required that the Proca fluctuations on such a geometry 
(in the probe and ultra-relativistic limits) do not experience negative time delays upon crossing the pp-wave\footnote{It is easy to see that the violation of 
null-energy condition may result in negative time
delays in high-energy scattering, i.e., superluminal propagation. This point has been duly noted already in~\cite{{Dubovsky:2005xd,Nicolis:2009qm,Rubakov:2014jja}}.}.
This further constrains the $(\a,\b)$ parameter plane to the region~(\ref{alphacons}).
Following from Eq.~(\ref{cond101}), a stronger bound on $\b$ would appear if we had required that null-energy condition continues to hold for distance scales smaller
than the effective field theory can resolve, i.e., $1/\L$. This would lead us to the conclusion: $\b<0$, which coincides with the findings reported
in~\cite{Jimenez:2013qsa}. However, as already noted by some authors~\cite{Drummond:1979pp,Goon:2016une} such extrapolations may be misleading.

As we have seen, the positivity constraints$-$derived from unitarity and analyticity of scattering amplitudes$-$have been quite inadequate since dominant contributions from
$t$-channel graviton exchange diagrams preclude any useful bounds on the effective field theory parameters. Nevertheless, it has been instructive to carry out this analysis;
it helps us better appreciate the power of the shock-wave analysis, which is particularly useful in constraining theories that involve massless particles. Surely, the two
analyses are inequivalent, and sometimes they give complementary results~\cite{Camanho:2016opx,Cheung:2016yqr}.
The shock-wave analysis captures $2\!\rightarrow\!2$ high-energy scattering events, or more precisely, resums horizontal ladder diagrams in the deflectionless limit:
$t/s\!\rightarrow\!0$, as noted in~\cite{Camanho:2014apa}.

It would be natural to extend the analyses presented in this article to more general Einstein-Proca
theories~\cite{Tasinato:2014eka,Heisenberg:2014rta,Hull:2015uwa,Allys:2015sht,Jimenez:2016isa,Heisenberg:2016eld}, whose parameter space may thereby be constrained.
Another interesting direction is to do similar studies in the presence of a (negative) cosmological constant, especially because of the availability of various
holographic techniques that could constrain bulk gravity theories (see for
example~\cite{Camanho:2014apa,Brigante:2007nu,Brigante:2008gz,Buchel:2009tt,Hofman:2009ug,Kulaxizi:2012xp,Kulaxizi:2015fza}
and references therein for a partial list of work in this direction). We leave this as future work.

\subsection*{Acknowledgments}
We are grateful to B.~Bellazzini, I.~L.~Buchbinder, L.~Heisenberg, K.~Hinterbichler, M.~Kulaxizi, J.~Maharana, K.~Mkrtchyan, M.~Porrati, S.~Theisen and A.~J.~Tolley
for valuable discussions and useful comments.

\begin{appendix}
\numberwithin{equation}{section}

\section{Appendix}\label{sec:appendix}

Throughout the bulk of the article, we have omitted some cumbersome expressions and technical details for the sake of readability.
The purpose of this appendix is to make room for those details. We begin with the Planck-suppressed graviton self interaction terms:
\begin{align}
\mathcal{L}_{\text{EH}}^{(\text{int})}&=\tfrac{2}{M_P}\left(\de_\m h_{\n\r}\de^\m h^{\r\s}h_\s{}^\n-\tfrac{1}{2}\de^\m h^{\r\s}\de^\n h_{\r\s}h_{\m\n}
-h_{\m\n}h_{\r\s}\de^\m\de^\n h^{\r\s}+h_{\m\r}h_{\n\s}\de^\m\de^\n h^{\r\s}\right)\nonumber\\
&-\tfrac{2}{M_P}\left(\de_\m h^{\m\r}\de_\n h^{\n\s}h_{\r\s}+\de_\m h'\de^\m h^{\r\s}h_{\r\s}+\tfrac{1}{2}\de_\m\de_\n h'h^{\m\r}h^\n{}_\r
-\de_\m h'\de_\r h^\r{}_\n h^{\m\n}\right)\label{St15}\\
&+\tfrac{1}{2M_P}h'\left(\de_\m h'\de^\m h'-\de_\m h^{\n\r}\de^\m h_{\n\r}+2h_{\m\n}\de^\m\de^\n h'+2\de_\m h^{\m\r}\de_\n h^\n{}_\r\right)
+\mathcal O(h^4),\nonumber\end{align}
where $h'\equiv h^\m_\m$\,. Next, we spell out the expansion of the $\sqrt{-g}g^{\m\n}$ term:
\beq \sqrt{-g}g^{\m\n}=\h^{\m\n}\!-\!\tfrac{2}{M_P}\left(h^{\m\n}-\tfrac{1}{2}\h^{\m\n}h'\right)+\tfrac{4}{M_P^2}\left(h^{\m\r}h_\r{}^\n-\tfrac{1}{2}h'h^{\m\n}-\tfrac{1}{4}
\h^{\m\n}\left(h_{\a\b}^2-\tfrac{1}{2}h'^2\right)\right)+\mathcal O(h^3).\eeq{St312a}
We also need an expansion for the Proca covariant derivative; it is given by:
\beq \nabla_\m B_\r=\de_\m B_\r+\tfrac{1}{M_P}\left(\h^{\a\b}-\tfrac{2}{M_P}h^{\a\b}+\cdots\right)\left(\de_\a h_{\m\r}-2\de_{(\m}h_{\r)\a}\right)B_\b,\eeq{St101}
where the ellipses stand for higher-order terms.

The $3\hspace{-0pt}\times\hspace{-0pt}3$ matrices $\mathcal{C}_{IJ}$ and $\mathcal{D}_{IJ}$ appearing in Eq.~(\ref{DE-0}) will appear below.
The matrix $\mathcal{C}$ is given by:
\beq \mathcal{C}_{IJ}=\left[\,\ln\!\left(\L b\right)-2\b\left(\L b\right)^{\!-2}\left(\d_{I2}-\d_{I3}\right)\,\right]\d_{IJ}
-\left[\,i(pb)^{\!-1}-2\b(q/p)\left(\L b\right)^{\!-2}\,\right]\d_{I2}\d_{J1},\eeq{C-matrix}
where we have \emph{no sum} over repeated indices. The matrix $\mathcal{D}$ takes the form:
\beq \mathcal{D}_{IJ}=\mathcal{D}_{IJ}^{(0)}+\b(\L b)^{\!-2}\,\mathcal{D}_{IJ}^{(1)}+\b^2(\L b)^{\!-4}\,\mathcal{D}_{IJ}^{(2)},\eeq{D-matrixS}
whose elements are specified below. With the short-hand notations: $k_\n\equiv K_\n(mb)=$ and $\e\equiv mb$, the non-zero
$\mathcal{O}\!\left(\b^0\right)$-components are given by:
\begin{equation}\label{D-matrix0}
\begin{split}
&\mathcal{D}_{11}^{(0)}=\left(1+\a\right)^2k_0^2+\a\left(1+\a\right)k_1^2,\qquad \mathcal{D}_{22}^{(0)}=\mathcal{D}_{33}^{(0)}=\left(1+\a\right)k_0^2,\\
&\mathcal{D}_{21}^{(0)}=\a\left(1+\a\right)(q/p)\left(k_0^2+k_1^2\right)+2i\left(1+\a\right)(m/p)\left[(1+2\a)k_0+\a k_1/\e\right]k_1,
\end{split}
\end{equation}
while the non-zero $\mathcal{O}\!\left(\b\right)$-components read:
\begin{equation}\label{D-matrix1}
\begin{split}
\mathcal{D}_{11}^{(1)}=&-\tfrac{1}{2}\a\left(\e k_0+k_1\right)k_1+\e^2\left[k_0^2-k_1^2+\tfrac{3}{8}\a\left(k_0^2-k_2^2\right)\right],\\
\mathcal{D}_{22}^{(1)}=&4\e(1+\a)k_0k_1+\e^2\left(k_0^2-k_1^2\right),\\
\mathcal{D}_{33}^{(1)}=&-4\e(1+\a)k_0k_1-4\e^2\left[\left(\a+\tfrac{3}{4}\right)k_0^2+\left(\a+\tfrac{5}{4}\right)k_1^2\right],\\
\mathcal{D}_{21}^{(1)}=&-\tfrac{1}{2}(q/p)\left[\a k_1^2+\e(9\a+8)k_0 k_1-\tfrac{3}{4}\e^2\a\left(k_0^2-k_2^2\right)\right]\\
&+i(m/p)\left[-11\a\left(k_0+k_1/\e\right)k_1-\tfrac{1}{4}\left(11\a k_0^2+8(1+\a)k_1^2-3\a k_2^2\right)\right],
\end{split}
\end{equation}
and the non-zero $\mathcal{O}\!\left(\b^2\right)$-components are:
\beq \mathcal{D}_{22}^{(2)}=-4\e^2k_1^2,\quad\mathcal{D}_{33}^{(2)}=6\e^2\left[k_1^2+\tfrac{1}{3}\e k_0k_1-\tfrac{1}{4}\e^2\left(k_0^2-k_2^2\right)\right],
\quad\mathcal{D}_{21}^{(2)}=4(q/p)\e^2k_1^2.\eeq{D-matrix2}
This marks the end of our short technical appendix.
\end{appendix}

\end{document}